\documentclass[a4paper,10pt]{article}

\newcommand{\bfa}[1]{\mbox{\boldmath${#1}$}}

\newcommand{\bea}{\begin{eqnarray}}
\newcommand{\eea}{\end{eqnarray}}
\newcommand{\bnn}{\begin{eqnarray*}}
\newcommand{\enn}{\end{eqnarray*}}
\newcommand{\be}{\begin{equation}}
\newcommand{\ee}{\end{equation}}

\def\PACS{\par\leavevmode\hbox {\it PACS:\ }}%
\def\MSC{\par\leavevmode\hbox {\it MSC:\ }}%
\def\UK{\par\leavevmode\hbox {\it Keywords:\ }}%
\begin{document}

\title{Symmetries in the hyperbolic Hilbert space}

\author{S. Ulrych\\ Wehrenbachhalde 35, CH-8053 Z\"urich, Switzerland}
\date{May 17, 2005}
\maketitle

\begin{abstract}
The $Cl(3,0)$ Clifford algebra is represented with the
commutative ring of hyperbolic numbers $H$. The canonical form of the Poincar\'e mass 
operator defined in this vector space corresponds to a sixteen-dimensional structure. This conflicts
with the natural perception of a four-dimensional space-time. The assumption that
the generalized mass operator is an hermitian observable forms the basis of a mathematical model that decomposes
the full sixteen-dimensional symmetry of the hyperbolic Hilbert space. The result is a direct product of the Lorentz group, the
four-dimensional space-time, and the hyperbolic unitary group $SU(4,H)$, which is
considered as the internal symmetry of the relativistic quantum state. The internal symmetry is
equivalent to the original form of the Pati-Salam model.
\end{abstract}

{\scriptsize\PACS{12.10.-g; 03.65.Fd; 11.30.Ly; 02.10.Hh; 02.10.De}
\MSC{81V22; 11E88; 30G35; 46E20; 15A33}
\UK{Hyperbolic complex Clifford algebra; Unified theories; Hyperbolic numbers; Paracomplex numbers; Split-complex numbers}}

\section{Introduction}
The complex numbers are naturally related to rotations and dilatations
in the plane, whereas the so-called hyperbolic numbers can be
related to Lorentz transformations and dilatations in the two-dimensional Minkowski
space-time. The hyperbolic numbers are also known as perplex \cite{Fje86}, unipodal \cite{Hes91}, duplex \cite{Kel94} or split-complex numbers.
Keller \cite{Kel94} has shown that the hyperbolic numbers, together with
the complex numbers and quaternions form the fundamental blocks employed in
the classification of Clifford algebras (see also Porteous \cite{Por81,Por95}).

Hucks investigated \cite{Huc93} that a four-component Dirac spinor is equivalent 
to a two-component hyperbolic spinor, and that the Lorentz group is equivalent to the hyperbolic unitary group
$SU(2,\bfa{H})$. In this work Hucks also found that the operations of C, P, and T on 
Dirac spinors are closely related to the three types of complex conjugation 
that exist when both hyperbolic and ordinary imaginary 
units are present.

Kocik \cite{Koc99} has shown that non-relativistic quantum mechanics over the hyperbolic numbers
behaves well in the probabilistic
interpretation via the Born's formula.
Generalizing Born's probability interpretation Khrennikov \cite{Khr03,Khr203} 
found that unitary transformations
in a hyperbolic Hilbert space represent a new class of transformations of probabilities
which describe a kind of hyperbolic interference.
The bicomplex Born formulas were investigated by Rochon and Tremblay \cite{Tre04}
for bicomplex wave functions. Note, that the
bicomplex numbers are isomorphic to the hyperbolic numbers, if they
are defined as an extension of the complex numbers. The hyperbolic numbers are a specialization 
of the hypercomplex numbers. An investigation of the geometries generated
by these numbers has been presented by Catoni et al. \cite{Cat05}.

It has been shown by Baylis and Jones \cite{Bay89} that a $Cl(3,0)$ Clifford algebra
is rich enough in structure to describe relativity as well as
the more usual $Cl(1,3)$ or $Cl(3,1)$. This approach is chosen also in this work,
where the algebra is represented with the hyperbolic numbers.

In a recent article \cite{Ulr05} it has been shown that the quadratic Dirac equation
and the Maxwell equations can be derived from the Poincar\'e mass operator
expressed in terms of a $Cl(3,0)$ Clifford algebra. However, a rigorous mathematical consideration of the used vector
space implies a sixteen-dimensional space-time structure, which contradicts to the
observation of three space dimensions and the intuitive observation of one time
dimension. The assumption that the generalized mass operator is hermitian forms the basis of a model
that decomposes the hyperbolic Hilbert space into a $SO(3,1,\bfa{R})\times SU(4,\bfa{H})$ symmetry.

The Lorentz group is naturally identified with four-dimensional space-time.
The $SU(4,\bfa{H})$ group is considered as the 
internal symmetry of the relativistic quantum state. The internal symmetry
is equivalent to the original form of the Pati-Salam model \cite{Pat73}. 
Pati and Salam proposed the group $SU(4,\bfa{C})\times SU(4,\bfa{C})$
to unify the fundamental hadrons and leptons into a common irreducible representation.
From this group they can generate a gauge theory of the weak, electromagnetic, and the strong 
interactions.

In \cite{Ulr05} a representation is used, which emphasizes the importance of the hyperbolic unit.
Here, the hyperbolic and complex units are considered as peers, therefore
the mathematical representation changes in some points, but the mathematical content remains
the same. 

\section{The hyperbolic vector space}
\label{space}
Vector spaces can be defined over the commutative ring of 
hyperbolic numbers $z\in\bfa{H}$.
\be
\label{beg}
z=x+iy+jv+ijw\;,\hspace{0.5cm}x,y,v,w \in\bfa{R}\;,
\ee
where the complex unit $i$ and the hyperbolic unit $j$ have the properties
\be
i^2=-1\;,\hspace{0.5cm}j^2=1\;.
\ee
The hyperbolic numbers are a commutative extension of the complex numbers to
include new roots of the polynomial equation $z^2 - 1 = 0$.
In the terminology of Clifford algebras they are represented by $Cl(1,0)$,
i.e. they correspond to a one-dimensional Clifford algebra.

A conjugation can be defined that reverses the sign of the complex and
the hyperbolic unit.
\be
\label{conj}
\bar{z}=x-iy-jv+ijw\;.
\ee 
With respect to this conjugation the square of the hyperbolic number can
be calculated as
\be
\label{square}
z\bar{z}=x^2+y^2-v^2-w^2+2ij(xw-yv)\;.
\ee

Relativistic vectors can be represented 
as linear combinations of the matrices $\sigma_\mu=(1,j\sigma_i)$,
where the elements of the Pauli algebra are multiplied by the hyperbolic unit.
A contravariant Lorentz vector then has the form
\be
\label{veco}
P=p^\mu\sigma_\mu\;.
\ee
The Clifford conjugation $\bar{P}$, the involution of spatial reversal, is represented
as the conjugation defined in Eq.~(\ref{conj}) plus a transposition of the basis
matrices.
Note, that the above algebra is congruent to the $Cl(3,0)$ paravector algebra of Baylis \cite{Bay89}.

\section{Hermiticity in the hyperbolic Hilbert space}
\label{Hilbert}
In \cite{Ulr05} the Poincar\'e mass operator has been introduced as 
$M^2=P\bar{P}$ in a four-dimensional space-time. It has been assumed that the
vector operators correspond to real coordinates $p^\mu\in\bfa{R}^{\,3,1}$ in momentum space.
However, from a mathematical point of view there is no reason, 
why the operator coordinates should be real.
In general they could be hyperbolic numbers. In the following the 
hyperbolic complexification $h^\mu\in\bfa{H}^{\,3,1}$ denotes the
hyperbolic momentum, 
which is isomorphic to a sixteen-dimensional real vector. This leads to a
generalized version of the mass operator
\be
\label{mass}
M^2=H\bar{H}\;,
\ee
with $H=h^\mu\sigma_\mu$.
However, this sixteen-component structure contradicts with our intuitive perception of a 
four-dimensional space-time world. To solve this problem a model is suggested,
which is based on the hermiticity of the mass operator in the hyperbolic Hilbert space.

It is one of the fundamental principles of quantum mechanics that observables
are represented by hermitian operators. In the representation of relativistic
quantum physics given here this means that an operator $A$ is equal to its conjugated operator
\be
\label{hermite}
A=\bar{A}\;,
\ee
and in addition that all eigenvalues with respect to the
eigenvectors of the operator are real numbers. Note, that this is a stronger conditon than
in non-relativistic quantum physics, because from Eq.~(\ref{conj}) it follows that a hyperbolic eigenvalue
with $a=\bar{a}$ is not by default a real number.

Since the mass operator represents an observable it is considered
to be hermitian even in its extended sixteen-dimensional form. Let $\vert\psi\rangle$
be an eigenstate of
the mass operator.
Then one can write based on the preceding arguments
\be
\label{cond}
M^2\,\vert \psi\rangle=m^2 \,\vert\psi\rangle\;,\hspace{0.5cm}m^2\in\bfa{R}\;.
\ee
In the following it is assumed that a state is not on the
light-cone. This
case needs a special treatment, which is
beyond the scope of this work. 

\section{The mass operator in vacuum}
\label{vac}
If the square defined in Eq.~(\ref{square}) is required to be a real number, 
a time-like hyperbolic number, which is not on the light-cone, can be
written in the form
\be
\label{hcoord}
h=p\,e^{i\phi+j\xi}\;,\hspace{0.5cm}p,\phi,\xi \in\bfa{R}\;.
\ee
The hyperbolic number $h=p\,\eta\,$ is decomposed into the magnitude, represented by a positive or negative real number, and 
a $U(1,\bfa{H})$  phase, which is normalized to unity with respect to the square. The phase is equivalent to
$U(1,\bfa{C})\times U(1,\bfa{C})$ and $SO(2,\bfa{R})\times SO(2,\bfa{R})$. Due to the normalization the phase
corresponds to two unit spheres in two-dimensional real Euklidean vector spaces, i.e. $S^1 \times S^1$. The geometrical structure is
visible in Eq.~(\ref{square}).

This principle is extended to the sixteen-dimensional case. 
A general hyperbolic vector can be written as
\be
\label{vecbeg}
Z=X+iY+jV+ijW\;,\hspace{0.5cm}x^\mu,y^\mu,v^\mu,w^\mu \in\bfa{R}^{\,3,1}\;.
\ee
The square of this vector is calculated as
\be
\label{vecsquare}
Z\bar{Z}= x_\mu x^\mu + y_\mu y^\mu - v_\mu v^\mu - w_\mu w^\mu +2ij(x_\mu w^\mu - y_\mu v^\mu )\;.
\ee

Consider again Eq.~(\ref{mass}).
To obtain a real square the hyperbolic momentum is decomposed 
into the four-dimensional magnitude, transforming with the Lorentz group, and a four-dimensional phase,
transforming with the hyperbolic unitary group.
In the rest frame a standard vector can be
chosen according to
\be
\label{state}
h^\mu\cong
\left(\begin{array}{c}
m\\
0\\
0\\
0\\
\end{array}\right)\!\times\!
\left(\begin{array}{c}
1\\
0\\
0\\
0\\
\end{array}\right)=
p^\mu \eta^i\;.
\ee
A general hyperbolic momentum vector is obtained with appropriate \linebreak$SO(3,1,\bfa{R})\!\times SU(4,\bfa{H})$ transformations
of the standard vector. 
The phase vector is normalized to unity in analogy to the one-dimensional case.
With these relationships one finds for the mass operator in vacuum
acting on a plane wave state $\vert \psi\rangle=\vert p^\mu\rangle\!\times\vert \eta^i\rangle$
\bea
\label{momform}
M^2\,\vert \psi\rangle= p_\mu p^\mu \,\vert\psi\rangle\;,\hspace{0.5cm}p_\mu p^\mu \in\bfa{R}\;.
\eea
Note, that in vacuum the mass operator reduces to $M^2= h_\mu \bar{h}^\mu$.
The phases can be transformed with the hyperbolic unitary group
without affecting the direction of the momentum and the expectation value of the mass operator.
The phase transformations are therefore understood as an internal symmetry.

The internal symmetry is equivalent to the symmetry groups
$SU(4,\bfa{C})\times SU(4,\bfa{C})$ and $SO(6,\bfa{R})\times SO(6,\bfa{R})$.
In the real representation one recovers, together with the
Lorentz group, a group acting in a sixteen-dimensional real vector space.
Due to the normalization $\eta^i$ corresponds to two unit spheres in six-dimensional real Euklidean vector spaces, i.e. 
$S^5 \times S^5$. 
The spheres are related to the twelve spatial coordinates $x^i,y^i$ and $v^i,w^i$ in Eqs.~(\ref{vecbeg})
and~(\ref{vecsquare}).

\section{Conclusions}
The origin of the proposed model can be found in Eq.~(\ref{conj}).
From this equation it follows that a hyperbolic number with the property $z=\bar{z}$
is not necessarily a real number. In addition, the square $z\bar{z}$ is not necessarily
real. Fundamental principles of quantum mechanics like hermiticity of observables and
Born's probability interpretation therefore have
a larger impact on the geometrical structure of the mathematical representation
than in non-relativistic quantum physics.

Based on the internal symmetry the gauge symmetries may be introduced. Note, that a gauge field changes
the structure of the mass operator \cite{Ulr05}. The new mass operator must be hermitian as well.
Current conservation could imply further boundary
conditions. This could break down the vacuum symmetry of the mass operator and might require the gauge symmetries
to be effectively defined in subgroups of $SU(4,\bfa{H})$.

\end{document}